\documentstyle[12pt]{article}
\def\be{\begin{equation}}
\def\ee{\end{equation}}
\def\bea{\begin{eqnarray}}
\def\eea{\end{eqnarray}}

\def\pd{\partial}

\def\L{\cal L}
\begin{document}
\title{Dirac-Born -Infeld Equations}
\author{ D.B. Fairlie$\footnote{e-mail: david.fairlie@durham.ac.uk}$\\
Department of Mathematical Sciences,\\
         Science Laboratories,\\
         University of Durham,\\
         Durham, DH1 3LE, England}
\maketitle
\begin{abstract}
Properties of the Dirac-Born-Infeld Lagrangian analogous to those of the 
Nambu-Goto String are analysed. In particular  the Lagrangian is shown to
be constant or zero on the space of solutions of the equations of motion if the Lagrangian is taken to any  power other than $\frac{1}{2}$.
\end{abstract}
\newpage
\section{Introduction}
It is well known that, at a classical level, the Schild Lagrangian for strings
\cite{schild}, which is the square of the Nambu-Goto Lagrangian which itself is a measure of the area swept out by the string, has equations of motion which are  a subset of those arising in the latter case. There is, however another constraint in the case of the Schild action; the Lagrangian is a constant or zero on the space of solutions. This property is not a requirement for the Nambu-Goto string, which therfore possesses a wider class of solutions.  The question arises as to whether there is something special about the two-dimensional nature of the world sheet, or whether Lagrangians, analogous to the Schild case can be constructed
in the case of the Dirac$\footnote{Historical remark: I was present at a seminar Dirac gave in Cambridge in 1962 on his extensible model of the electron. I well remember the atmosphere of indulgent scepticism with which the audience greeted his results!}$ -Born-Infeld Lagrangian, which pertains to the motion of
D-branes\cite{callan}\cite{dirac} \cite{born}. Indeed there is; The equations of motion for any function of the  Dirac-Born-Infeld Lagrangian will always be the same at the classical level. This property follows from the following result;
the gradient of the Lagrangian in  any direction within the world volume is 
zero on the space of solutions of the equations of motion for this Lagrangian
whenever it is taken to any power other than the square root.
 This is analogous to the statement for strings that longitudinal and timelike variations have no effect. In the special case of the square root Lagrangian
itself,
the conclusion that the gradient vanishes is evaded and the requirement that the Lagrangian should be a constant on the space of solutions is not mandatory.
The most significant consequence of the reparametrisation invariance is that the conserved stress tensor vanishes  as does that for the string, or Self Dual Yang Mills. 

 This characteristic of integrable systems, and suggests that the Born -Infeld theory may be at least partially tractable in any dimension. 
The Dirac-Born-Infeld Lagrangian ${\cal L}$ describes a mapping of the coordinates $x_j$ of a world volume of dimension $n$ embedded in a total space of dimension $d$, co-ordinatised by $X^\mu(x_j)$. An Abelian gauge field $\displaystyle{F_{i,j}=\frac{\partial A_j}{\partial x_i}\,-\,\frac{\partial A_i}{\partial x_j}}$ on the world volume is included
in the description, but additional terms of Chern-Simons type and supersymmetric terms are neglected.
The Lagrangian density is then
\be 
{\L}\,=\,\sqrt{\det{| g_{ij}+F_{ij} |}}\,=\, \sqrt{\det\left| \frac{\partial X^\mu}{\partial x_i}\frac{\partial X^\mu}{\partial x_j}\,+\,\left( \frac{\partial A_j}{\partial x_i}\,-\,\frac{\partial A_i}{\partial x_j}\right)\right|}\label{born}
\ee
The action from this Lagrangian is re-parametrisation invariant \cite{gibbons} with respect to transformations
of the world-volume co-ordinates $x'= x'(x_k)$, or in terms of differentials,
\be
 dx'_j\, =\, \frac{\partial x'j}{\partial x_k}dx_k \label{diff}\ee
 provided that the gauge fields transform as
\be
A'^j \,=\, A^k \frac{\pd x'_k}{\pd x_j}.\label{trans}
\ee
Then it is easy to see that
\be {\L}'\, =\,\left( \frac{\partial \{x'_1,x'_2,\dots x'_n\}}{\partial \{x_1,x_2,\dots x_n\}}\right){\L} \label{scale}
\ee
in other words ${\L}$ scales with the Jacobian of the transformation.
Note that this transformation is compatible with gauge transformations on $A^j$.
This invariance is also obviously compatible with the  gradient conditions which states that
\be
 \frac{\pd F({\L})}{\pd x_j}\,=\,0,\ \ j\,=\,1,\dots n,\label{compatible}
\ee
for any non trivial function of the Lagrangian.
In the next two section the proofs  of these assertions about the gradient law
and the stress tensor will be given. For simplicity, the metric on the world volume will be taken to be Euclidean, but this is not essential. Other recent approaches to the classical treatment of the Dirac-Born-Infeld Lagrangian may be found in references \cite{gibbons}\cite{deser}\cite {jackiw}. 
\section{Proofs of the gradient condition}
First of all, a reminder of how the result works for Strings. 

Let \(L(\sigma,\tau)\) be the Schild Lagrangian

\be L\,=\,
\det\left|\begin{array}{cc}
\frac{\partial X^\mu}{\partial \sigma}  \frac{\partial X^\mu}{\partial \sigma}&
\frac{\partial X^\mu}{\partial \sigma}  \frac{\partial X^\mu}{\partial \tau}\\
\frac{\partial X^\mu}{\partial \tau}  \frac{\partial X^\mu}{\partial \sigma}&
\frac{\partial X^\mu}{\partial \tau}  \frac{\partial X^\mu}{\partial \tau}
\end{array}\right|\label{schild}
\ee
We want to show that the equations of motion from \({\L}=\sqrt{L}\) are included  classically in those from \(L\)

Consider the equation of motion
\be\frac{\partial}{\partial\sigma}\frac{\partial L}{\partial  \frac{\partial X^\mu}{\partial \sigma}}+\frac{\partial}{\partial\tau}\frac{\partial L}{\partial  \frac{\partial X^\mu}{\partial \tau}}=0\label{eqmo}
\ee
Now multiply this by \( \frac{\partial X^\mu}{\partial \sigma}\) and sum over 
\(\mu\). We shall show that this is equivalent to \( \frac{\partial L}{\partial \sigma}\), which therefore vanishes. (  \( \frac{\partial L}{\partial \sigma}\), by a slight abuse of notation , means the derivative of $L$ with respect to $\sigma$ holding $\tau$ fixed).
Consider the equation

\begin{equation}
\frac{\partial X^\mu}{\partial \sigma}\left(\frac{\partial}{\partial \sigma}\frac{\partial L}{\partial  \frac{\partial X^\mu}{\partial \sigma}}+\frac{\partial}{\partial \tau}\frac{\partial L}{\partial  \frac{\partial X^\mu}{\partial \tau}}\right)=0\label{identity}
\end{equation}
The left hand side is equal to
\be
\frac{\partial }{\partial \sigma}\left(\frac{\partial X^\mu}{\partial \sigma}\frac{\partial L}{\partial  \frac{\partial X^\mu}{\partial \sigma}}\right)+\frac{\partial }{\partial \tau}\left(\frac{\partial X^\mu}{\partial \sigma}\frac{\partial L}{\partial  \frac{\partial X^\mu}{\partial \tau}}\right)-\frac{\partial^2X^\mu }{\partial\sigma^2}\frac{\partial L}{\partial  \frac{\partial X^\mu}{\partial \sigma}}-\frac{\partial^2X^\mu }{\partial\sigma\partial\tau}\frac{\partial L}{\partial  \frac{\partial X^\mu}{\partial \tau}}\ee
 But \(L\) is homogeneous of degree 2 in \( \frac{\partial X^\mu}{\partial \sigma}\)  and \( \frac{\partial X^\mu}{\partial \tau}\). The term\hfill\break
\(\displaystyle{\frac{\partial }{\partial \sigma}\left(\frac{\partial X^\mu}{\partial \sigma}\frac{\partial L}{\partial  \frac{\partial X^\mu}{\partial \sigma}}\right)}\) reduces to \(\displaystyle{2\frac{\partial L}{\partial \sigma}}\) and the term \( \displaystyle{\frac{\partial }{\partial \tau}\left(\frac{\partial X^\mu}{\partial \sigma}\frac{\partial L}{\partial  \frac{\partial X^\mu}{\partial \tau}}\right)\,=\,0}\), by the theorem of false co-factors for determinants. The last two terms reduce to  \(\displaystyle{\frac{\partial L}{\partial \sigma}}\) so the final result is
that the equation (\ref{identity}) and the similar one for $\tau$ derivatives reduce to
\be \frac{\partial L}{\partial \sigma}\,=\, \frac{\partial L}{\partial \tau}\,=\,0\label{gradienteq}\ee
Now if instead, the equations of motion  are sought
 which follow from  \(L^{p}\), taking as Lagrangian an arbitrary power, we obtain
\be
\frac{\partial}{\partial \sigma}pL^{p-1}\left(\frac{\partial L}{\partial  \frac{\partial X^\mu}{\partial \sigma}}\right)+\frac{\partial}{\partial \tau}pL^{p-1}\left(\frac{\partial L}{\partial  \frac{\partial X^\mu}{\partial \tau}}\right)\,=\,0\label{identity2}.
\ee
A similar manipulation to that above, by multiplying (\ref{identity2}) by 
$\frac{\pd X^\mu}{\pd \sigma}$ and summing over $\mu$ yields the equation
\be (2p-1)\frac{\partial L}{\partial \sigma}\,=\,0,\label{gradient2eq}\ee
so either $p\,=\,\frac{1}{2}$, or else $\frac{\partial L}{\partial \sigma}\,=\,0.$
Now expanding  (\ref{identity2}), we obtain
\be
L\left(\frac{\partial}{\partial \sigma}\left(\frac{\partial L}{\partial  \frac{\partial X^\mu}{\partial \sigma}}\right)+\frac{\partial}{\partial \tau}\left(\frac{\partial L}{\partial  \frac{\partial X^\mu}{\partial \tau}}\right)\right)+(p-1)\left(\frac{\partial L}{\partial  \frac{\partial X^\mu}{\partial \sigma}}\frac{\partial L}{\partial \sigma}+\frac{\partial L}{\partial  \frac{\partial X^\mu}{\partial \tau}}\frac{\partial L}{\partial \tau}\right)\,=\,0\label{identity3}
\ee
and the last two terms vanish on account of (\ref{gradienteq}) when the 
equations of motion for \(L\) are satisfied.
Hence the equations of motion from \(L^p\)  are satisfied whenever the equations of motion from  \(L\) are for any $p$.  However, in the case of the square root Lagrangian ${\L}$ the deduction that $\frac{\pd {\L}}{\pd \sigma}$ and $\frac{\pd {\L}}{\pd \tau}$ both vanish and that hence ${\L}$ is necessarily constant, or zero when the equations of motion are satisfied cannot be drawn. 
This argument clearly also works in higher dimensions.The key result is that \({\bf n}\cdot\nabla L\) vanishes for all vectors \(\bf n\) lying in the tangent space to the world surface whenever the equations of motion for $L$ are satisfied.
The  inclusion of an Abelian field $F_{ij}$ into the Lagrangian 
 makes no essential difference to this argument. The proof follows along similar lines, as follows. Take now 
\be {L}\,=\,{{\det}|g_{ij}+F_{ij}|}\,=\, {{\det}\left| \frac{\partial X^\mu}{\partial x_i}\frac{\partial X^\mu}{\partial x_j}\,+\,\left( \frac{\partial A_j}{\partial x_i}\,-\,\frac{\partial A_i}{\partial x_j}\right)\right|}\label{born2}
\ee
the square of (\ref{born}). 
 Let \( \hat L_{ij}\) denote the cofactor (i.e. signed minor) of the \(i, j\) th element of the matrix  with determinant \(L\). Then it is straightforward to see that
\begin{equation}
\frac{\partial L}{\partial  \frac{\partial X^\mu}{\partial x_k}}\,=\,\sum_j
\frac{\partial X^\mu}{\partial x_j}(\hat L_{jk}\,+\,\hat L_{kj})\label{diff1}
\end{equation}
and
\begin{equation}
 \frac{\partial L}{\partial  \frac{\partial A^j}{\partial x_k}}\,=\,\hat L_{jk}\,-\,\hat L_{kj}\label{diff2}
\end{equation}

The proof of this result follows the lines of the proof when $F_{\mu\nu}$ is absent.

It works by combining (\ref{diff1}) with (\ref{diff2}) and using the result
\begin{equation}
\sum_j
\left(\frac{\partial X^\mu}{\partial x_i}\frac{\partial X^\mu}{\partial x_j}+F_{ij}\right)\hat L_{jk}\,+\,\left(\frac{\partial X^\mu}{\partial x_i}\frac{\partial X^\mu}{\partial x_j}-F_{ij}\right)\hat L_{kj}\,=\, 2L\delta_{ik}
\label{falseco}
\end{equation}
which summarises a set of determinantal identities sometimes known as the theorem of false co-factors. In virtue of the equations of motion we have the identity
\begin{equation}
\sum_j^n\frac{\partial X^\mu}{\partial x_i}\left(\frac{\partial}{\partial x_j}\frac{\partial L}{\partial  \frac{\partial X^\mu}{\partial x_j}}\right)\,+\,\sum_{j}\sum_{k\neq i}F_{ik}\left(\frac{\partial}{\partial x_j }\frac{\partial L}{\partial  \frac{\partial A^k}{\partial x_j}}\right)\,=\,0\label{identit2}
\end{equation}
Pulling out the derivative with respect to \(x_j\) as before and using (\ref{diff1}) and (\ref{diff2})  we see that this type of relation can be re-expressed as 
\be\frac{\partial L}{\partial x_i}\, =\, 0.\label{cond}
\ee 
The term 
\be\sum_j\sum_k -\frac{\partial^2 A^i}{\partial x_j\partial x_k}\frac{\partial L}{\partial  \frac{\partial A^k}{\partial x_j}}\,\equiv\,0,\ee
as a consequence of the fact that \(\displaystyle {\frac{\partial L}{\partial  \frac{\partial A^k}{\partial x_j}}}\) is antisymmetric in \(j,\ k\).
Thus the derivative of $L$ with respect to any of the world volume co-ordinates vanishes on the space of solutions to the equations of motion It is easy to see that these conditions imply that the same equation of motion results from 
taking any differentiable function of $L$ as a Lagrangian.
\section{$L$ as the sum of quadratic terms}
All these higher dimensional Lagrangians share a common property with the Lagrangian for a free particle; namely that $L$, given by (\ref{born2}) can be expressed as the sum of quadratic terms. The easiest way to explain this result is to consider the case where the dimensions of the world volume dimensions $n,\,=\,4$.
Then $L$ is the determinant
\be
\det\left|\begin{array}{cccc}
\frac{\partial X^\mu}{\partial x_1}  \frac{\partial X^\mu}{\partial x_1}&
\frac{\partial X^\mu}{\partial x_1}  \frac{\partial X^\mu}{\partial x_2}+F_{12}&
\frac{\partial X^\mu}{\partial x_1}  \frac{\partial X^\mu}{\partial x_3}+F_{13}&
\frac{\partial X^\mu}{\partial x_1}  \frac{\partial X^\mu}{\partial x_4}
+F_{14}\\
\frac{\partial X^\mu}{\partial x_2}  \frac{\partial X^\mu}{\partial x_1}
+F_{21}&
\frac{\partial X^\mu}{\partial x_2}  \frac{\partial X^\mu}{\partial x_2}&
\frac{\partial X^\mu}{\partial x_2}  \frac{\partial X^\mu}{\partial x_3}
+F_{23}&
\frac{\partial X^\mu}{\partial x_2}  \frac{\partial X^\mu}{\partial x_4}
+F_{24}\\
\frac{\partial X^\mu}{\partial x_3}  \frac{\partial X^\mu}{\partial x_1}
+F_{31}&
\frac{\partial X^\mu}{\partial x_3}  \frac{\partial X^\mu}{\partial x_2}
+F_{32}&
\frac{\partial X^\mu}{\partial x_3}  \frac{\partial X^\mu}{\partial x_3}&
\frac{\partial X^\mu}{\partial x_3}  \frac{\partial X^\mu}{\partial x_4}
+F_{34}\\
\frac{\partial X^\mu}{\partial x_4}  \frac{\partial X^\mu}{\partial x_1}
+F_{41}&
\frac{\partial X^\mu}{\partial x_4}  \frac{\partial X^\mu}{\partial x_2}
+F_{42}&
\frac{\partial X^\mu}{\partial x_4}  \frac{\partial X^\mu}{\partial x_3}
+F_{43}&
\frac{\partial X^\mu}{\partial x_4}  \frac{\partial X^\mu}{\partial x_4}
\end{array}\right|
\ee
This may be expressed as

\bea 
L&=&\frac{1}{4!}\sum_{\mu\nu\rho\sigma} \left(\epsilon_{ijkl}\frac{\partial X^\mu}{\partial x_i}  \frac{\partial X^\nu}{\partial x_j}\frac{\partial X^\rho}{\partial x_k}  \frac{\partial X^\sigma}{\partial x_l}\right)^2\nonumber\\
&+& \frac{1}{2!}\sum_{\mu\nu} \left(\frac{1}{2}\epsilon_{ijkl}\frac{\partial X^\mu}{\partial x_i}  \frac{\partial X^\nu}{\partial x_j}F_{kl}\right)^2
\,+\, \left(\frac{1}{4}\epsilon_{ijkl}F_{ij}F_{kl}\right)^2\label{quadrat}
\eea
This expression shows the pattern; the individual terms may be expressed as the contraction of an epsilon symbol with factors of the form $\frac{\partial X^\mu}{\partial x_i}$ ,or else   $F_{ij}$, where now the derivative acts on everything to the right, taking into account all possible terms. It is the analogue of the relativistic particle Lagrangian $\sqrt{\frac{\partial X^\mu}{\partial \tau}\frac{\partial X^\mu}{\partial \tau}}$. This expression may be written more succinctly as
\be
L\,=\, \sum_M \Phi^M\Phi^M,\label{short}
\ee
where the important property of the determinants $\Phi^M$ is that they are all
expressible as divergences. This means that the equations of motion for $X^\mu$
(and similarly for $A^i$) take the form
 \be\sum_M\sum_i\left(\frac{\pd\Phi^M}{\pd x_i}\right)\left(\frac{\partial\Phi^M}{\partial  \frac{\partial X^\mu}{\partial x_i}}\right)\,=\,0.\label{simple}
\ee

\section{The Stress Tensor.}
Another remarkable property of the Lagrangian ${\L}$ is that the associated stress tensor automatically vanishes.
A gauge invariant stress tensor,  $T_{ij}$ for  $L$
is given by
\bea
T_{ij}&=& \sum_{\mu=1}^d\frac{\pd X^\mu}{\pd x_i}\frac{\partial L}{\partial  \frac{\partial X^\mu}{\partial x_j}}+\sum_{k=1}^nF_{ik}\frac{\partial L}{\partial  \frac{\partial A^k}{\partial x_j}}-L\delta_{ij}\nonumber\\
&=& \sum_{\mu=1}^d \frac{\pd X^\mu}{\pd x_i},\sum_j
\frac{\partial X^\mu}{\partial x_k}(\hat L_{jk}\,+\,\hat L_{kj})\,+\,\sum_{k=1}^nF_{ik}(\hat L_{jk}\,-\,\hat L_{kj})-L\delta_{ij}\label{stress}\\
&=& L\delta_{ij}.\nonumber
\eea
The last step follows from the determinantal identities. This tensor is purely diagonal, and divergencless as a consequence of the conditions (\ref{cond}).
When the Lagrangian ${\L}\,=\,\sqrt{L}$ is used the terms above cancel and even the diagonal components vanish! This is a consequence of the reparametrisation invariance of the theory. 

\section{Discussion}
The three properties of the Dirac-Born-Infeld Lagrangian, diffeomorphism covariance, vanishing gradients on the world volume for generic Lagrangians and vanishing stress tensor
are obviously inter-related. The vanishing of the stress tensor is in fact a consequence of reparametrisation invariance. These properties are, of course,  well known in the special case of the Nambu-Goto string. This fact, together with the circumstance that the stress tensor for Yang Mills in 4-dimensions vanishes on instanton solutions, strongly suggests that the Dirac-Born-Infeld theory may be more tractable than has been hitherto suspected. One may ask why  the gradient property has apparently not been noticed before. A possible answer is that in most treatments the induced  metric $\displaystyle {g_{ij}\,=\, \frac{\pd X^\mu}{\pd x_i}\frac{\pd X^\mu}{\pd x_j}}$ is split into contributions from the world volume, by setting $X^i =x^i$ together with contributions from the transverse space, {\it before} evaluating the equations of motion, i.e.
\be
g_{ij} \,=\,\eta_{ij}\,+\,\sum_{\mu=n+1}^{\mu=d} \frac{\pd X^\mu}{\pd x_i}\frac{\pd X^\mu}{\pd x_j}. \label{metric}
\ee
The derivation given here more closely parallels the standard treatment of the string, where the co-ordinate choice is not made until after the equations of motion are derived.
An interesting  mathematical question arises from this work. It is well known,
\cite{fai} that the Schild action may be regarded as the $N\rightarrow\infty$
limit of Yang Mills in strong coupling with gauge group $SU(N)$. What is the corrresponding finite $N$ version of the Dirac-Born-Infeld theory?
\section*{Acknowledgements}
I am indebted to Linda Baker, Peter Bowcock, Holger Nielsen and Cosmas Zachos 
for useful discussions.

\end{document}